\documentclass{elsart}
\usepackage{graphicx}
\usepackage{amsmath}
\usepackage{amssymb}

\journal{Journal of Crystal Growth}

\begin{document}

\begin{frontmatter}

\title{Differential thermal analysis and solution growth of intermetallic compounds}

\author[ALCMP,ALMEP]{Y. Janssen\corauthref{cor}}
\corauth[cor]{Corresponding author.} \ead{yjanssen@ameslab.gov}
\author[ALCMP]{M. Angst}
\author[ALMEP]{K.W. Dennis}
\author[ALMEP]{R.W. McCallum}
\author[ALCMP,ISU]{P.C. Canfield}

\address[ALCMP]{Condensed Matter Physics, Ames Laboratory, 50011 Ames IA}
\address[ALMEP]{Materials and Engineering Physics, Ames Laboratory, 50011 Ames IA}
\address[ISU]{Department of Physics and Astronomy, Iowa State University, 50011 Ames IA}
\begin{abstract}

To obtain single crystals by solution growth, an exposed primary
solidification surface in the appropriate, but often unknown,
equilibrium alloy phase diagram is required. Furthermore, an
appropriate crucible material is needed, necessary to hold the
molten alloy during growth, without being attacked by it.
Recently, we have used the comparison of realistic simulations
with experimental differential thermal analysis (DTA) curves to
address both these problems. We have found: 1) complex DTA curves
can be interpreted to determine an appropriate heat treatment and
starting composition for solution growth, without having to
determine the underlying phase diagrams in detail. 2) DTA can
facilitate identification of appropriate crucible materials. DTA
can thus be used to make the procedure to obtain single crystals
of a desired phase by solution growth more efficient. We will use
some of the systems for which we have recently obtained
single-crystalline samples using the combination of DTA and
solution growth as examples. These systems are TbAl,
Pr$_7$Ni$_2$Si$_5$, and YMn$_4$Al$_8$.

\end{abstract}


\end{frontmatter}

\section{Introduction}

Solution growth emulates a way in which nature often produces
single crystals of minerals, i.e.\ out of a liquid with a
composition that is different from the
product~\cite{Fisk89,Canfield92,Canfield01}. In our laboratory, as
well as in others, many of the materials that are produced for
investigations of their physical properties, are grown as single
crystals by solution growth. Often, when it comes to producing
single crystals of a desired phase, insufficient phase-diagram
data is available, and we must estimate which composition and
temperature ranges may produce the desired phase as well as what
crucible to use. Then, based on the products of such experiments,
it is decided if and how to alter the the initial composition,
temperature range and crucible material. This process can, at
times, require multiple iterations that consume both time and
resources.

Recently, we have expanded the use of  differential thermal
analysis (DTA) as part of this procedure. We have found that the
use of DTA can greatly facilitate the optimization of the growth,
and can play a role in selecting the right crucible material. In
this paper, which is focussed on solution growth, the comparison
of realistic simulations with experimental differential thermal
analysis (DTA) curves is used to obtain growth parameters without
detailed knowledge of phase diagrams, whereas prior use of DTA for
crystal growth mainly involved detailed and lengthy phase-diagram
studies (see e.g. Ref.~\cite{Schultze91}).

In the following, after describing the experimental techniques, we
will present simulations of DTA signals for a hypothetical binary
system. Then we will discuss the DTA-assisted growth of three
compounds that may serve as examples: TbAl, Pr$_7$Ni$_2$Si$_5$,
and YMn$_4$Al$_8$. Of these phases, mm-sized single crystals have
not been produced before.

The descriptions of solution growth in this paper are by no means
complete. Rather, with its focus is on the use of DTA for
determining solution-growth parameters, it should be considered an
extension to earlier papers, by Fisk and Remeika~\cite{Fisk89},
Canfield and Fisk~\cite{Canfield92} and Canfield and
Fisher~\cite{Canfield01}.

\section{Experimental}

Differential thermal analysis (DTA) was performed in a PerkinElmer
Pyris DTA 7 differential thermal analyzer. As a process gas, we
used Zr-gettered ultra-high-purity Ar. For crucibles we used
Al$_2$O$_3$ (manufactured by PerkinElmer), MgO (custom-made by
Ozark Technical Ceramics, Inc.), and Ta (home-made from
small-diameter Ta tubes). In order to protect the Pt cups and the
thermocouple of our instrument from possible Ta diffusion, the Ta
crucibles were placed inside standard ceramic crucibles. The
samples, with mass $\sim$40-80 mg, were made by arc melting
appropriate amounts of starting materials with typical (elemental)
purities of 99.9-99.99\,\%. An experiment consisted of two or
three cycles at heating and cooling rates of typically
10-40$^\circ$C/min.
The data from the first heating cycle was different from data from
subsequent heating cycles. This may have occurred because the
sample shape did not conform to the crucible so that it was not in
intimate contact with the crucible walls until it had melted, or
because a reaction with the crucible changed the composition of
the sample. In a DTA curve, besides the events described in
Sec.~3, there is also a baseline, not associated with the
properties of the sample (see e.g. Ref.~\cite{Wilburn58}). This
baseline is also influenced by the rate at which the DTA-unit
ramped.

For the growth experiments we used the following
procedure~\cite{Fisk89,Canfield92,Canfield01}. Appropriate amounts
of starting materials with typical (elemental) purities of
99.9-99.99\,\% were selected, pre-alloyed by arc melting, if
needed, and put into a crucible. The crucible material was the
same as for the DTA experiment, Al$_2$O$_3$ (Coors), MgO (Ozark
Tech.), or Ta (homemade). For separating the grown crystals from
the remaining liquid, a sieve was used. For the ceramic crucibles,
an inverted crucible catches the liquid, while a plug of quartz
wool in the catch crucible acted as a sieve. The Ta crucibles were
`3-cap crucibles'~\cite{Canfield01}, with a built-in sieve. The
crucibles were placed in a flat-bottom quartz ampoule with some
quartz wool above and below the crucibles, to prevent possible
cracking caused by differential thermal expansion between the
quartz and the crucible and to provide cushioning during the
decanting process. The quartz ampoule was evacuated and filled
with a partial pressure of Ar, so that the pressure in the ampoule
was nearly atmospheric at the highest temperature, and was then
sealed. The ampoule was then placed in a box furnace, and subject
to a heat treatment determined from DTA experiments (see Sec.~3).
After the final temperature had been reached, the ampoule was
taken out of the furnace, inverted into the cup of a centrifuge
and quickly spun to decant the liquid from the crystals.

For initial characterization, we measured powder X-ray diffraction
patterns on one or several finely ground crystals from the growth
yield with a Rigaku Miniflex+ diffractometer employing
Cu-K$\alpha$ radiation. The patterns were analyzed with
Rietica~\cite{Rietica}, using a Le Bail-type~\cite{LeBail88} of
refinement.

\section{Simulations}

For the evaluation of measured DTA curves, we compared them to
simulated, ideal, DTA curves. For the simulations, we considered a
hypothetical binary system, the $\alpha$-$\delta$ system (Fig.~1).
This system is composed of $\alpha$ and $\delta$ that melt
congruently at 700$^\circ$C and 1000$^\circ$C or 1400$^\circ$C,
respectively, peritectic compounds $\beta$ and $\gamma$, at
$\alpha_{50}\delta_{50}$ and $\alpha_{25}\delta_{75}$,
respectively, and a eutectic alloy of composition
$\alpha_{15}\delta_{85}$ with a eutectic temperature of
600$^\circ$C. At its decomposition temperature, 800$^\circ$C,
peritectic $\beta$ decomposes into a liquid with composition
$\alpha_{60}\delta_{40}$ and solid $\gamma$, which in turn
decomposes, at 900$^\circ$C, into a liquid with composition
$\alpha_{40}\delta_{60}$ and solid $\delta$. For the liquidus
lines we chose second-order polynomials with the concentration of
$\delta$ as a variable. This functional dependence can be
considered realistic~\cite{Okamoto91}, and is easily evaluated.

\begin{figure}[!tb]
\begin{center}
\includegraphics[angle=270,width=0.6\textwidth]{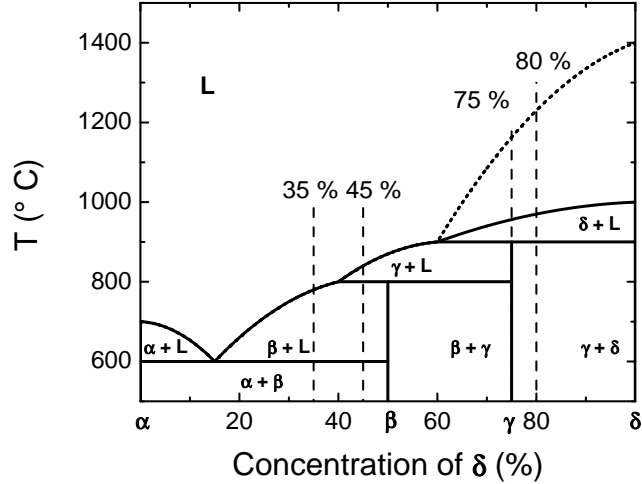}
\caption{Hypotetical binary phase diagrams containing two
congruently melting phases, $\alpha$ and $\delta$ and two
peritectics, $\beta$ and $\gamma$. The dotted line is for $\delta$
melting at 1400$^\circ$C. The dashed vertical lines indicate
compositions for which DTA curves were calculated.}\label{HypPD}
\end{center}
\end{figure}

A full discussion, see e.g.\ Ref.~\cite{Boettinger02}, of DTA is
beyond the scope of this paper and we shall limit our
consideration to a rather simple model. We will consider the DTA
as a black box, that produces a signal proportional to the
temperature derivative of the enthalpy of the sample. We assume
that the enthalpies of formation for the phases $\alpha$, $\beta$,
$\gamma$, and $\delta$ are equal, and that the specific heats for
the solid phases and the liquid (regardless of composition), are
constant and equal. Then the simulated DTA signal is proportional
to the temperature derivative of the fraction of solid to liquid,
determined by the well-known lever law (see e.g.\
Ref.~\cite{Verhoeven75}). Our DTA measurements were usually
carried out at heating and cooling rates of 10 to 40$^\circ$C/min,
which does not usually result in an equilibrium distribution of
phases. Therefore, a cooling curve was calculated assuming that
once a phase has solidified it does not dissolve back, or react
with other solid phases. Then the composition of the liquid
follows the liquidus line and the final solid contains a
nonequilibrium distribution of phases. The calculated heating
curve represents the heating curve of this nonequilibrium
distribution of phases under the assumption that the kinetics are
such that also during heating the liquidus line is followed.
Although the assumptions are not fully realistic, the patterns
recognizable in experimentally observed curves are reproduced well
by our model.

During a DTA measurement (even in the absence of undercooling),
there is a difference between the events observed in the heating
and cooling curves. The melting and solidification events have the
same onset temperature, but show a width (dependent on the heating
or cooling rate). To include this effect in our simulations as an
instrument response function, we assumed this temperature lag is
temperature independent and can be described by the `standard
$\Gamma$ distribution', that is often used to describe waiting
times~\cite{Hogg66}.

\begin{figure}[!tb]
\begin{center}
\includegraphics[width=0.6\textwidth]{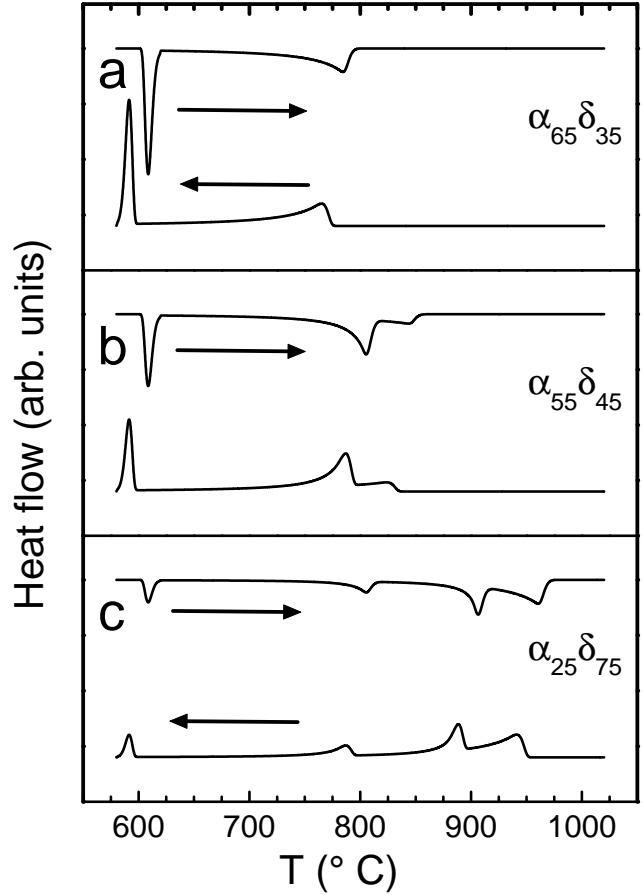}
\caption{Calculated DTA curves for various compositions of the
phase diagram in Fig.~1. In each example, the lower curve is the
cooling curve for a non-equilibrated sample, and the top curve is
the heating curve for a non-equilibrated sample. }\label{DTAsims}
\end{center}
\end{figure}

In Fig.~2a the calculated DTA curves for an alloy of composition
$\alpha_{65}\delta_{35}$ are presented (see dashed vertical line
in Fig.~1). Events at both the eutectic (600$^\circ$C) and
liquidus (780$^\circ$C) temperatures are clearly observed. The
eutectic is a much more sharply defined event than the liquidus,
because the observed thermal event is proportional to the
temperature derivative of the fraction of solid-to-liquid. In
Fig.~2b, the calculated DTA curves for an alloy of composition
$\alpha_{55}\delta_{45}$ are shown (see dashed vertical line in
Fig.~1). The eutectic (600$^\circ$C), and the peritectic
decomposition of $\beta$ (800$^\circ$C) are clearly observed in
the calculations, whereas the liquidus (840$^\circ$C) is less
obvious. In Fig.~2c, the calculated DTA curves for an alloy of
composition $\alpha_{25}\delta_{75}$ are shown for the case that
$\delta$ melts at 1000$^\circ$C (see dashed vertical line in
Fig.~1, crossing the lower-lying liquidus line). The eutectic
(600$^\circ$C), the peritectic temperature of $\beta$
(800$^\circ$C), the peritectic temperature of $\gamma$
(900$^\circ$C) and the liquidus (956$^\circ$C) are visible. Note
that in the simulations for Figs.~2c the peritectic temperatures
for the $\beta$-phase and the eutectic are visible because of the
non-equilibrium nature of the model.

Such DTA curves can be used, especially in the case that the
underlying phase diagram is unknown, to determine the temperature
range over which crystals can be grown. For example, the DTA
curves for a sample of composition $\alpha_{65}\delta_{35}$ show
that the primary solidification for that composition can be grown
by slowly cooling between 800$^\circ$C and 600$^\circ$C. In
addition, to separate the crystals from the liquid, the sample
should be decanted above 600$^\circ$C. After the growth, the
crystals will be identified as $\beta$, e.g.\ by X-ray
diffraction. Crystals of the $\gamma$-phase form for
$\alpha_{55}\delta_{45}$ and can be grown by slowly cooling
between the liquidus, 850$^\circ$C, and 800$^\circ$C, and
separated by decanting above 800$^\circ$C. However, choosing the
composition $\alpha_{55}\delta_{45}$ is quite demanding: the
maximum useful temperature range for growth is only about
30-50$^\circ$C. Moreover, the weight fraction of crystals to
liquid will be low for this composition. Finally, crystals of the
$\delta$-phase form for $\alpha_{25}\delta_{75}$ and can be grown
(for the lower-lying liquidus line in Fig.~1) by cooling slowly
between 970$^\circ$C, and separated by decanting above
900$^\circ$C.

\begin{figure}[!tb]
\begin{center}
\includegraphics[angle=270,width=0.6\textwidth]{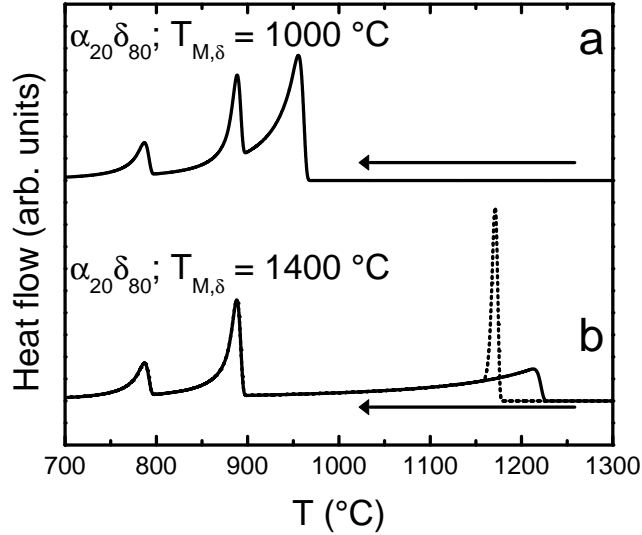}
\caption{Simulated DTA cooling curves for $\alpha_{20}\delta_{80}$
of the phase diagram in Fig.~1, for ({\bf a}) $\delta$ melting at
1000$^\circ$C and ({\bf b}) at 1400$^\circ$C. The dotted line in
({\bf b}) represents a cooling curve with undercooling by
50$^\circ$C taken into account.}\label{DTA1400}
\end{center}
\end{figure}

The simulations indicate that the event associated with the
liquidus can be quite weak (see Figs.~2a and b). As we mentioned
above, the strength of the thermal signature for a liquidus is
highly dependent on the slope of the liquidus curve, and can be
very hard to detect experimentally. Consider an identical phase
diagram, except that $\delta$ melts at 1400$^\circ$C (dotted in
Fig.~1). As a result the liquidus curve above 900$^\circ$C is
considerably steeper. In Fig.~3, modelled DTA cooling curves for
$\alpha_{20}\beta_{80}$ alloys from both hypothetical systems are
presented. As can be seen, the DTA signature for the system with a
lower-melting $\delta$ and a lower slope is much easier to detect,
i.e.\ removing more solid from the liquid per $^\circ$C yields a
stronger DTA signal.

Under the assumption that there is no time lag due to the
dissolution of the last solid phase, the calculated heating curves
can be considered realistic representations of actual
measurements. However, experimental cooling curves may be shifted
to lower temperatures due to undercooling. In many systems, the
nucleation of the crystalline phase from the liquid can be slow.
Since the DTA measurements were made at reasonably high cooling
rates this may lead to significant undercooling. (note that in an
actual growth experiment, typically at much slower cooling rates,
there is probably less undercooling.) In an undercooled melt, when
nucleation does occur, there is a rapid growth of the solid phase.
In a simulation for the composition $\alpha_{20}\delta_{80}$ with
$\delta$ melting at 1400$^\circ$C, allowing undercooling by
50$^\circ$C, the solidification is very clearly visible in the DTA
curve (Fig.~3b, dotted line). Thus whereas we may not detect the
true liquidus temperature we may be able to determine both an
upper bound given by the maximum temperature to which the sample
was heated, and a lower bound given by the solidification event
for the liquidus. We can give the upper bound, because when not
all crystals have been dissolved (i.e.\ the liquidus temperature
has not been reached) these serve as perfect nucleation centers,
preventing undercooling. Note that we have frequently observed
that peritectic temperatures and occasionally the eutectic
temperature are also undercooled.

\section{Growth of TbAl}

The growth of TbAl provides an example of the importance of
identifying the right crucible material, and the role DTA
measurements can play in this selection. Furthermore, it also
provides an example of the efficient determination of a very
narrow temperature range, by DTA, over which crystals can be
grown.

The crystal structure and magnetic properties, that involve
anisotropic exchange interactions, of polycrystalline TbAl were
published some time ago~\cite{Becle67,Barbara68,Becle68,Becle70}.
For an estimate of a composition for the initial melt for solution
growth, we considered the binary phase diagram of the Tb-Al
system. To our knowledge, this has only been
predicted~\cite{Ferro93,Ferro94}, from the systematics of
different rare-earth - Al binary systems, but not verified
experimentally in detail. Especially the Tb-rich side (above
$\sim$~56\,\% Tb) is uncertain. This predicted phase diagram
includes 2 eutectics, at ~3.5\,\% (642$^\circ$C) and at ~77\,\% Tb
(903$^\circ$C), and 5 compounds: TbAl$_3$, TbAl$_2$, TbAl,
Tb$_3$Al$_2$, and Tb$_2$Al. Of these, TbAl$_2$ melts congruently,
whereas TbAl$_3$ and TbAl form peritectically. Even though
Tb$_3$Al$_2$ and Tb$_2$Al are indicated to form peritectically as
well, congruent melting cannot be excluded due to a lack of
experimental data~\cite{Ferro93}. Since the phase diagram around
this composition was indicated to be uncertain, a DTA experiment,
rather than a slow and expensive growth served as an inexpensive
and quick check, using only small amounts of starting materials.
The primary-solidification line for TbAl reportedly lies between
57\,\% and 67\,\% Tb and 1079-986$^\circ$C. Therefore, for the
experiments described below, we chose to use an alloy with
composition Tb$_{60}$Al$_{40}$.

In our experience, an alloy with more than 10-15\,\% rare earth
cannot be reliably held in an Al$_2$O$_3$ crucible, because of
thermite-type reactions. Furthermore, we considered it possible
that the Tb$_{60}$Al$_{40}$ alloy, being also rich in Al, would
attack Ta, the other crucible material we regularly used in the
past~\cite{Fisk89,Canfield92,Canfield01}. MgO, on the other hand,
could be stable enough to use. To test the crucibles, we performed
DTA experiments in the three available crucible materials:
Al$_2$O$_3$, Ta, and MgO. In order to prevent direct contact of
unreacted elements with the crucibles, we alloyed samples of
approximately 40 mg by arc-melting prior to the DTA experiments.
DTA heating and cooling cycles were performed three times between
$\sim$~1240 and $\sim$~600$^\circ$C, at heating rates of
40$^\circ$C/min and cooling rates of 10$^\circ$C/min. The curves
obtained from the first two cycles, between 800~$^\circ$C and the
highest temperature reached, are shown in Fig.~4.

\begin{figure}[!tb]
\begin{center}
\includegraphics[width=0.6\textwidth]{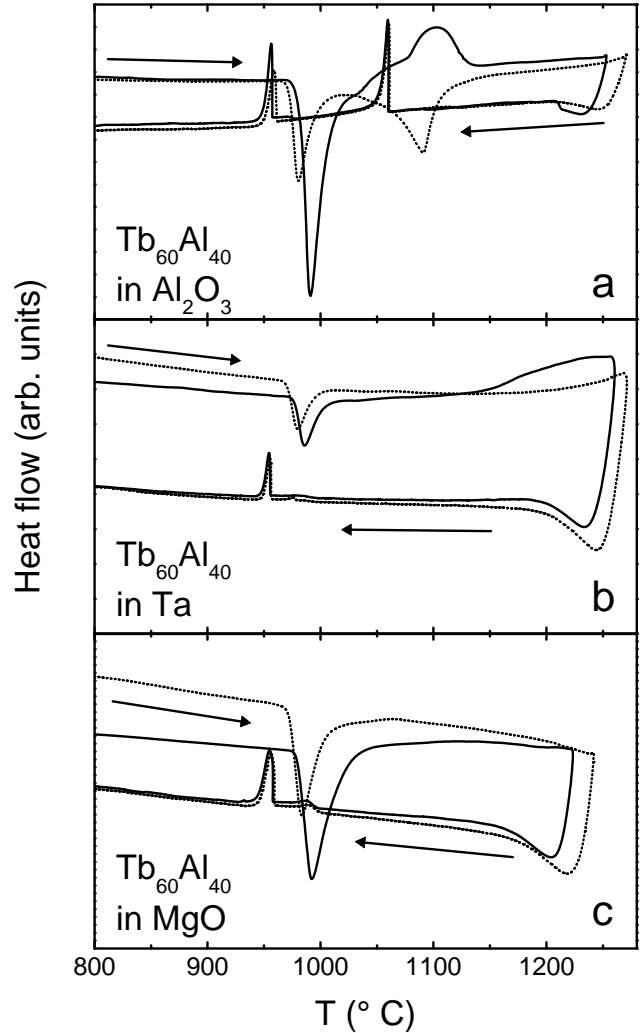}
\caption{The first (solid) and second (dotted) heating-and-cooling
cycles of Tb$_{60}$Al$_{40}$ in (a) Al$_2$O$_3$,(b) Ta, and (c)
MgO, measured upon heating with a 40$^\circ$C/min rate and cooling
with a 10$^\circ$C/min rate.} \label{TbAlDTAf}
\end{center}
\end{figure}

Upon heating, in both the first and the second heating cycle, an
endothermic event occurred near 970$^\circ$C in all three cases.
Besides a shift of the baseline, probably because the contact
between the sample and the thermocouple changed after melting, the
MgO and Ta-crucible curves show no significant difference between
the first and second heating curves, and no clear events at higher
temperatures. The two Al$_2$O$_3$-crucible heating curves,
however, are very different: a clear exothermic bump between
~1070$^\circ$C and ~1120$^\circ$C is visible in the first heating
curve, but not in the second one. Furthermore, the endothermic
event near 970$^\circ$C is less pronounced in the second heating
curve, and followed by a pronounced second endothermic event that
peaks near 1100$^\circ$C.

Upon cooling, (note that the non-linear but smooth behavior at
elevated temperatures in the cooling curves are due to
stabilization of the furnace-ramp rate, rather than due to thermal
events) the MgO and Ta-crucible curves show no significant
difference between the first and the second cooling cycle. In the
Al$_2$O$_3$-crucible curves, there are several differences between
the first and second cooling curves. Particularly, near
1210$^\circ$C, a weak exothermic event may be observed in the
first cooling curve, but not in the second cooling curve. In both
first and second cooling cycles, pronounced exothermic peaks were
observed near 1060$^\circ$C and 960$^\circ$C. Note that the peak
near 1060$^\circ$C is \emph{only} observed in the
Al$_2$O$_3$-crucible curves.

These results are consistent with a thermite-type reaction of the
alloy Tb$_{60}$Al$_{40}$ with the Al$_2$O$_3$ crucible, reducing
the amount of Tb in the metallic liquid, and increasing the amount
of Al. The exothermic bump in the first heating cycle, between
1080 and 1120$^\circ$C (Fig.~4a) then is associated with this
reaction. The event that occurred upon cooling at ~1060$^\circ$C,
is likely due to the peritectic temperature of TbAl, predicted to
be at 1079$^\circ$C~\cite{Ferro93,Ferro94}. Furthermore, the
highest-temperature event in the first cooling cycle was no longer
present in the second cooling cycle, maybe due to the sample
becoming so rich in Al that it could no longer fully melt (the
nearest compound richer in Al, Al$_2$Tb reportedly melts at
1514$^\circ$C).

Since both Ta and MgO showed no evidence of a reaction, both are
good candidates as a crucible material. (Although we worried that
the Tb$_{60}$Al$_{40}$ alloy would attack Ta, there is no clear
indication of such an attack from the DTA results.) For the growth
experiment, we chose MgO as the crucible material.

\begin{figure}[!tb]
\begin{center}
\includegraphics[angle=270,width=0.6\textwidth]{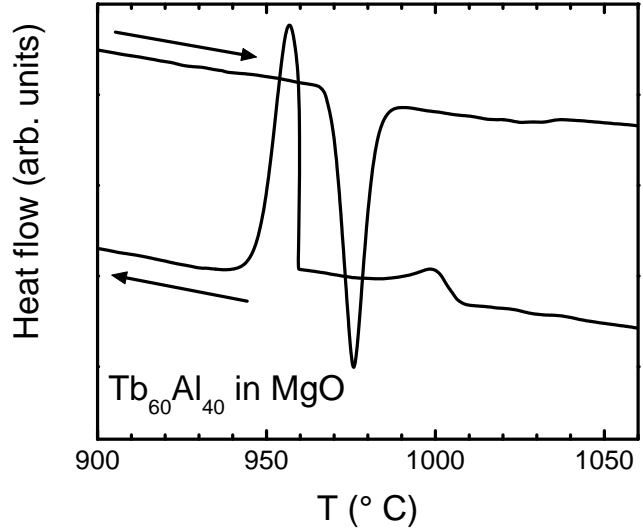}
\caption{Relevant part of the DTA curve of our sample of
Tb$_{60}$Al$_{40}$ measured upon heating with a 10$^\circ$C/min
rate and cooling with a 10$^\circ$C/min rate.}\label{DTATbAl}
\end{center}
\end{figure}

The heat treatment for the growth experiment is determined from
the details of the DTA curves. In Fig.~5, the relevant parts of a
third heating and cooling curve (measured at 10$^\circ$C/min for
both heating and cooling) are shown for an MgO-crucible
experiment. Upon heating, a sharp endothermic peak is observed
with an onset temperature of about 965$^\circ$C. This is followed
by a weak step near 1015-1035$^\circ$C. Upon cooling, two events
are observed. There is a weak exothermic bump with an onset
temperature of about 1010$^\circ$C, followed by a strong
exothermic peak with an onset temperature of about 960$^\circ$C.
Note the resemblance between these experimental results and the
simulation in Fig.~2a, and also note that there is some
undercooling of the liquidus, $\sim$5-30$^\circ$C, that makes the
liquidus much more visible upon cooling, as in the simulation in
Fig.~3. The measurements suggest that crystals of the primarily
solidifying compound for the composition Tb$_{60}$Al$_{40}$ can be
grown by cooling slowly between $\sim$1020$^\circ$C and
$\sim$965$^\circ$C, followed by a decant above $\sim$965$^\circ$C.
Taking into account the possibility of undercooling, this heat
treatment is very demanding (limited temperature range for
growth), and may be influenced by differences in thermometry
between the growth furnace and the DTA.

For the growth experiment, an arc-melted button ($\sim$3~g) was
first heated to 1200$^\circ$C, kept at this temperature for 2~h
for homogenization, then it was cooled in $\sim$1 h to
1020$^\circ$C, near the observed liquidus in the heating curve of
the DTA experiment. Finally, it was cooled in 10~h to
975$^\circ$C, which is near the temperature of the onset of the
endothermic event in the heating curve of the DTA experiment. At
this temperature, the sample was decanted. In the growth crucible,
large crystals of several mm were found (see the photograph in
Fig.~6a). Powder-X-ray diffraction identifies the crystals as
TbAl, with space group \emph{Pbcm}, and lattice parameters
a=5.85(3) \AA, b=11.4(3) \AA, and c=5.63(3) \AA, in agreement with
the reported crystal structure~\cite{Becle67}.

\begin{figure}[!tb]
\begin{center}
\begin{tabular}{ccc}
\includegraphics[angle=90,width=\textwidth]{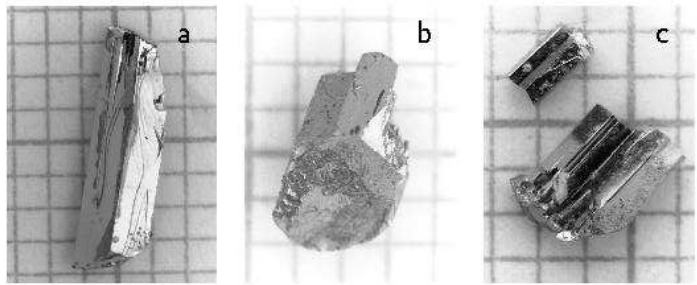}
\end{tabular}\caption{Photographs of the crystals of which the growth is described in the text.
    From left to right: {\bf(a)} TbAl, {\bf(b)} Pr$_7$Ni$_2$Si$_5$, {\bf(c)} YMn$_4$Al$_8$.
    In each case a mm grid was used as a background.}\label{photos}
\end{center}
\end{figure}

Crystals of TbAl could probably have been obtained from the alloy
Tb$_{60}$Al$_{40}$ based upon the available binary phase diagram
alone, cooling slowly below the reported liquidus temperature
($\sim$~1060$^\circ$C) and decanting above the reported peritectic
temperature for Tb$_3$Al$_2$ (986$^\circ$C). However, we have
often observed that binary phase diagrams with rare-earth
components need revision (see e.g.~\cite{Huang05}), and the Tb-Al
binary phase diagram was already known to be
uncertain~\cite{Ferro93,Ferro94}.

For the DTA simulations, we have assumed that the solidification
of an alloy follows the liquidus. Under these assumptions, it
seems from our DTA experiments on Tb$_{60}$Al$_{40}$, that the
solidification of TbAl is immediately followed by a last,
eutectic, solidification at a temperature substantially higher
than the eutectic temperature reported (903$^\circ$C). If the
alloy indeed followed the liquidus, the published binary phase
diagram~\cite{Ferro93} requires re-examination. However, further
experiments, which are outside the scope of this paper, would be
required to clarify this.

\section{Growth of Pr$_7$Ni$_2$Si$_5$}

We are currently investigating the ternary Pr-Ni-Si system,
including the liquidus surface of the Pr-rich corner.
Approximately 20 ternary intermetallic compounds have been
reported~\cite{Huang05} for the Pr-Ni-Si system, in accordance
with what was reported for Ce-Ni-Si~\cite{Rogl84}, and
Nd-Ni-Si~\cite{Bodak97}. One of those is the compound
Pr$_7$Ni$_2$Si$_5$.

For many Pr-Ni-Si compounds, the range of compositions that form
the primary-solidification surface is very narrow, sometimes down
to a few percent~\cite{Huang05b}. Therefore, it is useful to
quickly and  systematically test different compositions via DTA
analysis. In this section, we show that our DTA and growth
experiments, on an an alloy of composition
Pr$_{50}$Ni$_{25}$Si$_{25}$, resulted in the identification and
optimized growth of the compound that solidifies primarily:
Pr$_7$Ni$_2$Si$_5$.

A rod of several grams of composition Pr$_{50}$Ni$_{25}$Si$_{25}$
was prepared by arc-melting and drop-casting. Since the rod had
cooled down very quickly, we considered it homogeneous on the
scale of the DTA samples. Therefore, a piece of this rod, of the
typical size for a DTA experiment or a growth, was considered
representative for the whole rod. Two pieces were taken, about 40
mg for the DTA experiment, and about 2.25 g for the growth
experiment.

For the experiments, we used Ta crucibles as experience showed us
that, whereas Ta may be expected to be attacked by Ni and
Si~\cite{deBoer88}, these alloys (with about 50\,\% of Pr) do not
appear to attack the crucible.

\begin{figure}[!tb]
\begin{center}
\includegraphics[angle=270,width=0.6\textwidth]{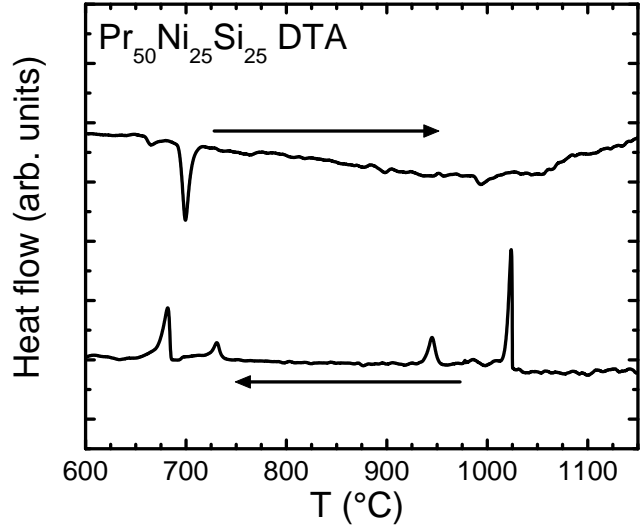}
\caption{Relevant part of the DTA curve of our sample of
Pr$_{50}$Ni$_{25}$Si$_{25}$ measured upon heating and cooling with
a 10$^\circ$C/min rate.}\label{DTAPrNiSi}
\end{center}
\end{figure}

For the DTA experiment, the sample was heated and cooled three
times between $\sim$400 and $\sim$1200$^\circ$C at
10$^\circ$C/min. After the first heating cycle, during which the
sample settled in the crucible, the measurements were
reproducible. In Fig.~7, the relevant parts of the third-cycle
curves are shown. The curves are substantially noisier than those
shown for the TbAl experiment. This may be because the thermal
contact between the Ta crucible and its ceramic liner varied, or
because the mass of the Ta crucible was substantially greater than
that of the sample inside it. In spite of the noise, we were able
to extract information, useful for crystal growth, from the
experiment.

In the cooling curve, four sharp exothermic events can be
observed, with onset temperatures of about 1025$^\circ$C,
950$^\circ$C, 730$^\circ$C, and 685$^\circ$C. This experimental
cooling curve can roughly be compared to the simulation of the
hypothetical binary alloy $\alpha_{25}\delta_{75}$, see Fig.~2c.
In that simulation, upon cooling, four events occur, the liquidus,
two events associated with peritectics, and the eutectic. The
experimental liquidus is probably sharpened by undercooling, see
Fig.~3. In the experimental heating curve, only one event with an
onset temperature of about 690$^\circ$C, can be observed clearly.
At higher temperatures, there may be an event near 990$^\circ$C (a
weak peak), and between $\sim$1050-1090$^\circ$ (a broad step).

Note that, whereas we do not fully understand the differences
between these heating and cooling curves (particularly in the
lower temperature range), for crystal growth we only need to know
the temperature region over which only the primary solidification
grows, i.e.\ between the liquidus temperature and the temperature
where secondary phases may start to grow.

The heat treatment for crystal growth might be proposed based upon
the DTA cooling curve alone, but we used the heating curve for
some guidance. For estimation of the liquidus, we examined at the
highest-temperature events.  In the cooling curve, the shape of
the peak near 1025$^\circ$C appears consistent with undercooling
(c.f.\ Fig.~3b). In the heating curve, the broad step between
$\sim$1050-1090$^\circ$ may be associated with that peak and
appears similar to the simulated liquidus in Fig.~2b. In the
cooling curve, the second-highest temperature event appears as a
sharp peak near 940$^\circ$ suggesting a decanting temperature
higher than 940$^\circ$ (but lower than 1025$^\circ$). The weak
peak in the heating curve at 990$^\circ$ may suggest that
secondary phases can start to grow below that temperature. Since
this temperature falls between the two highest-temperature events
in the cooling curve, we considered it safe to decant at a
temperature slightly above 990$^\circ$. Therefore, we used the
following heat treatment for the growth experiment. The sample was
heated to 1190$^\circ$C in 5 h, and allowed to equilibrate for 2
h. After that, it was cooled to 1100$^\circ$C in 2 h. After this
it was cooled to 1000$^\circ$C in 50 h, after which the sample was
decanted. In the crucible, large mm-sized blocky crystals were
found. A photograph of one of the crystals is displayed in
Fig.~6b.

Powder-X-ray diffraction identified the crystals as
Pr$_7$Ni$_2$Si$_5$, with space group is \emph{Pnma}, and lattice
parameters a=23.32(3) \AA, b=4.302(3) \AA, and c=13.84(3) \AA, in
agreement with the reported crystal structure~\cite{Myskiv74}. The
compound is only know by its crystal structure, therefore we are
currently investigating its low-temperature thermodynamic and
transport properties~\cite{Janssen04}. The combined DTA and
crystal-growth experiments demonstrate that composition
Pr$_{50}$Ni$_{25}$Si$_{25}$ is part of the primary phase field of
the compound Pr$_7$Ni$_2$Si$_5$.

\section{Growth of YMn$_4$Al$_8$ (and YMn$_2$Al$_{10}$)}

The growth of YMn$_4$Al$_8$ provides an example of how DTA can
help make finding the right composition for solution growth very
efficient. The ternary intermetallic compound YMn$_4$Al$_8$ has
been known at least since 1971 ~\cite{Rykhal71}, but has only been
synthesized in polycrystalline form. Recently~\cite{Nakamura04},
it was reported that YMn$_4$Al$_8$ has a narrow pseudogap in the
spin excitation spectrum. This prompted us to try to grow single
crystals.

The published partial triangulated ternary isotherm for the
Y-Mn-Al system~\cite{Rykhal71} at 600$^\circ$C suggests that a
growth may be attempted from an Al-rich liquid. Therefore, we
tried to grow it using a composition approximately halfway between
Al and YMn$_4$Al$_8$: Y$_{36}$Mn$_{145}$Al$_{819}$. We chose an
Al$_2$O$_3$ crucible, since in our experience such an alloy will
likely not attack it, and we started with pieces of the elements.
The sample was cooled in 15 h between $\sim$1200$^\circ$C and
950$^\circ$C, then decanted. The mm-sized prismatic crystals in
the growth crucible were identified as YMn$_2$Al$_{10}$, which was
not reported in the ternary isotherm, but does appear in
literature~\cite{Thiede98}.

As a next step, we alloyed by arc-melting a total of about 1 g of
the starting elements in the YMn$_4$Al$_8$ ratio. Although we had
some losses due to evaporation of Mn, about 2\,\%, after arc
melting, an X-ray powder diffraction experiment indicated the
sample to be mainly YMn$_4$Al$_8$. Then we performed DTA, in an
Al$_2$O$_3$ crucible, on a $\sim$20 mg piece. Up to 1350$^\circ$C
there was only one noticeable thermal event in both the heating
and cooling curves, at temperatures between 1220-1240$^\circ$C .
This, combined with the diffraction, is an indication that
YMn$_4$Al$_8$ is congruently melting at 1220-1240$^\circ$C.

A congruently melting material is part of its own primary
solidification surface. Therefore, we decided to try a composition
nearby YMn$_4$Al$_8$. The melting temperature of the desired phase
is higher than the reported liquidus temperature for an alloy of
composition Mn$_4$Al$_8$, $\sim$1100$^\circ$C~\cite{Jansson92},
therefore it seemed possible to grow YMn$_4$Al$_8$ out of a
composition close to the binary Mn-Al-line.

\begin{figure}[!tb]
\begin{center}
\includegraphics[angle=270,width=0.6\textwidth]{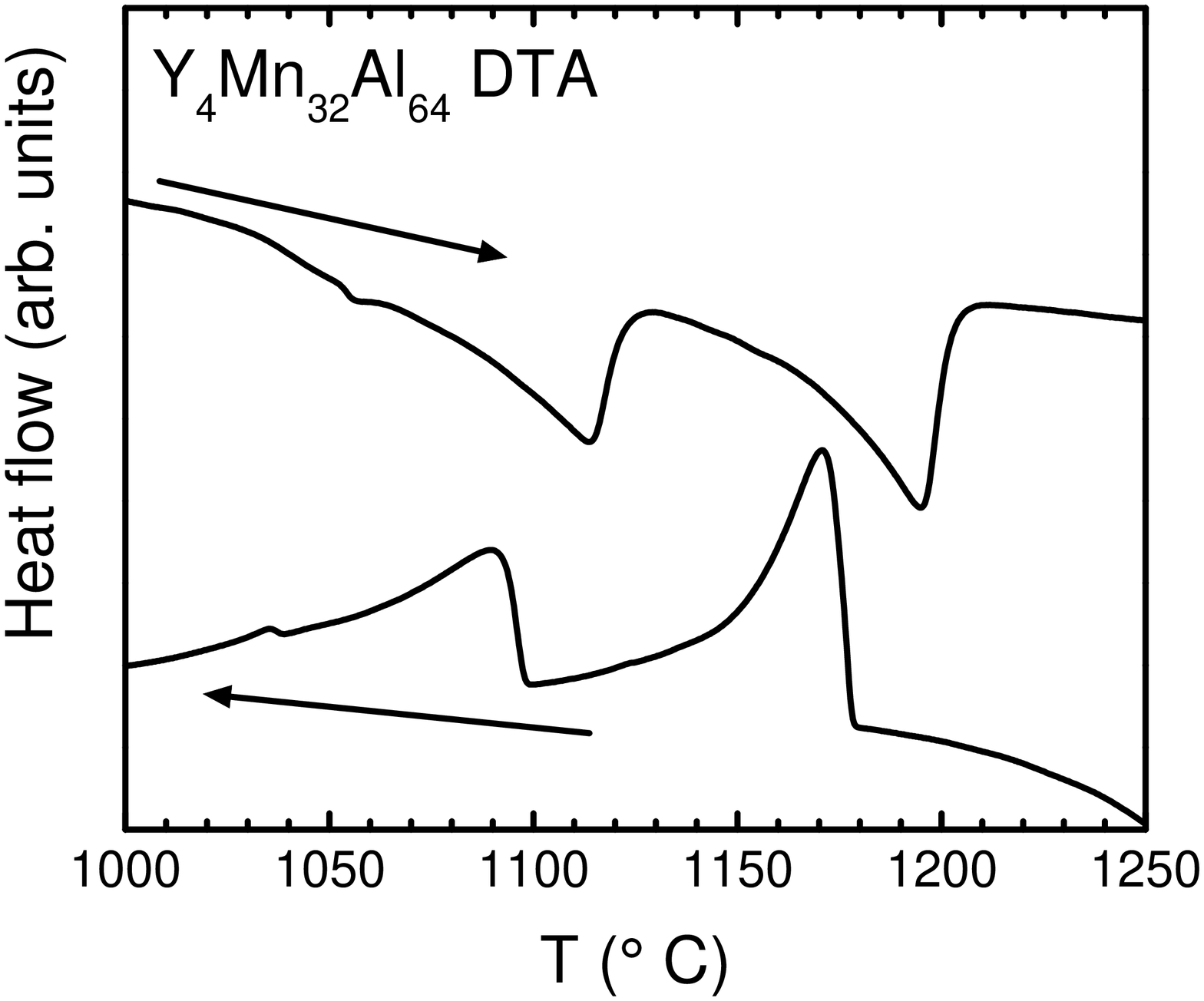}
\caption{Relevant part of the DTA curve of our sample of
Y$_{4}$Mn$_{32}$Al$_{64}$ measured upon heating and cooling with a
20$^\circ$C/min rate.}\label{DTAYMnAl}
\end{center}
\end{figure}

We decided to try an alloy of composition Y$_4$Mn$_{32}$Al$_{64}$
using Al$_2$O$_3$ crucibles. We alloyed a few grams by
arc-melting, and again had losses (about 3\,\%) due to Mn
evaporation. In order to make a small sample representative for
the whole, the arc-melted button was coarsely ground and the
powder thoroughly mixed. About 20 mg of the powder was used for a
DTA experiment. The results for heating and cooling at
20$^\circ$C/min, between 1000 and 1250$^\circ$C, are shown in
Fig.~8. In both the heating and the cooling curve, two very
pronounced events are visible. In the heating curve, peaks are
seen near 1110$^\circ$C and 1200$^\circ$C, while in the cooling
curve events occur at onset temperatures of $\sim$1180$^\circ$C
and $\sim$1100$^\circ$C. A weak event is observed at lower
temperatures of 1030-1050$^\circ$C in both the heating and cooling
curve. For the determination of growth parameters only the two
highest-temperature events are important, therefore we did not
measure down to still lower temperatures. These experimental
curves can be compared to the simulated curves in Fig.~2c.

The DTA experiment suggests that crystals can be grown and
separated from the remaining melt by cooling an alloy of
composition Y$_4$Mn$_{32}$Al$_{64}$ slowly below
$\sim$1200$^\circ$C (above the highest-temperature peak in the
heating curve) and decanting above $\sim$1130$^\circ$C (above the
second-highest-temperature peak in the heating curve). For a
growth experiment, we started with appropriate amounts of pieces
of the elements. The sample was first heated to 1250$^\circ$C for
equilibration, and then cooled in 1 h to 1200$^\circ$C, below
which is was cooled to 1160$^\circ$C in 60 h. At this temperature
the sample was decanted. Well-separated prismatic crystals were
found in the growth crucible. A photograph of two of those
crystals is presented in Fig.~6c. Powder-X-ray diffraction
identified the crystals as YMn$_4$Al$_8$, with space group
\emph{I4mmm}, and lattice parameters a=8.86(1) \AA, c=5.12(1) \AA,
in agreement with the reported crystal structure~\cite{Buschow76}.

\section{Summary and conclusions}

The examples presented here address how DTA can help in
determining growth parameters for solution growth, without
detailed knowledge of phase diagrams. As shown with the example of
TbAl, DTA can sometimes help in identifying the right crucible
material, and, moreover, it can help pinpointing a very narrow
temperature range over which to grow crystals. The example of the
growth of Pr$_7$Ni$_2$Si$_5$ shows that the combination of DTA and
growth experiments can help in determining the primarily
solidifying compound out of a given metallic liquid, while
limiting the growth to that of the primary. Finally, the example
of YMn$_4$Al$_8$ shows how DTA can help in the quick determination
of the primary phase field for a compound.

Extensions of the method can be sought in including other crucible
materials for DTA, e.g.\ BN. Or, in order to reduce problems with
elements that have high vapor pressures, in sealing the Ta-DTA
crucibles. However, problems still exist with elements that have a
high vapor pressure and cannot be held in Ta.

As was already discussed by Fisk and Remeika~\cite{Fisk89}, one of
the great advantages of solution growth is the economy of the
method. By including DTA in the procedure to optimize growth, it
can be economized even further, especially in terms of material
costs. Furthermore, although DTA generally shows that ``something
occurs at a certain temperature'' \cite{Schultze91}, the
combination of a DTA experiment with a growth experiment, if
successful, can lead to definite conclusions regarding the
primarily solidifying compound out of a metallic liquid of certain
composition. For this it is not necessary to establish a full
phase diagram, a DTA experiment on a sample of the composition of
interest is sufficient.

\section{Acknowledgements}

The authors wish to thank J. Fredericks, S. Chen, B. K. Cho, M.
Huang, D. Wu, T. A. Lograsso, S. L. Bud'ko, G. Lapertot for their
kind help in discussing and preparing samples. The financial
support from the US Department of Energy is gratefully
acknowledged: Ames Laboratory is operated for the US Department of
Energy by Iowa State University under Contract No. W-7405-Eng-82.
This work was supported by the Director for Energy Research,
Office of Basic Energy Sciences.


\begin{thebibliography}{10}
\expandafter\ifx\csname url\endcsname\relax
  \def\url#1{\texttt{#1}}\fi
\expandafter\ifx\csname
urlprefix\endcsname\relax\def\urlprefix{URL }\fi

\bibitem{Fisk89}
Z.~Fisk, J.~P. Remeika, in: {K. A. Gscheidner, Jr.}, L.~Eyring
(Eds.), Handbook
  on the Physics and Chemistry of Rare Earths, Vol.~12, Elsevier, Amsterdam,
  1989.

\bibitem{Canfield92}
P.~C. Canfield, Z.~Fisk, Philos. Mag. B 65 (1992) 1117.

\bibitem{Canfield01}
P.~C. Canfield, I.~R. Fisher, J. Cryst. Growth 225 (2001) 155.

\bibitem{Schultze91}
D.~Schultze, Thermochim. Acta 190 (1991) 77.

\bibitem{Wilburn58}
F.~W. Wilburn, J. Sci. Instrum. 35 (1958) 403.

\bibitem{Rietica}
B.~Hunter, Lhpm-rietica, www.rietica.org.

\bibitem{LeBail88}
A.~{Le Bail}, H.~Duroy, J.~L. Fourquet, Mater. Res. Bull. 23
(1988) 447.

\bibitem{Okamoto91}
H.~Okamoto, T.~B. Massalski, J. Phase Equilib. 12 (1991) 148.

\bibitem{Boettinger02}
W.~J. Boettinger, U.~R. Kattner, Metall. Mater. Trans. A 33A
(2002) 1779.

\bibitem{Verhoeven75}
J.~D. Verhoeven, Fundamentals of Physical Metallurgy, J. Wiley and
Sons, New
  York, 1975.

\bibitem{Hogg66}
R.~V. Hogg, A.~Craig, Introduction to Mathematical Statistics,
Macmillan, New
  York, 1966.

\bibitem{Becle67}
C.~B\`{e}cle, R.~Lemaire, Acta Crystallogr. 23 (1967) 840.

\bibitem{Barbara68}
B.~Barbara, C.~B{\`e}cle, R.~Lemaire, R.~Pauthenet, J. Appl. Phys.
39 (1968)
  1084.

\bibitem{Becle68}
C.~B\`{e}cle, R.~Lemaire, E.~Parthe, Solid State Commun. 6 (1968)
115.

\bibitem{Becle70}
C.~B\`{e}cle, R.~Lemaire, D.~Paccard, J. Appl. Phys. 41 (1970)
855.

\bibitem{Ferro93}
R.~Ferro, S.~Delfino, G.~Borzone, A.~Saccone, G.~Cacciamani, J.
Phase Equilib.
  14 (1993) 273.

\bibitem{Ferro94}
R.~Ferro, S.~Delfino, G.~Borzone, A.~Saccone, G.~Cacciamani, J.
Phase Equilib.
  15 (1994) 125.

\bibitem{Huang05}
M.~Huang, T.~A. Lograsso, J. Alloys Compd. 395 (2005) 75.

\bibitem{Rogl84}
P.~Rogl, in: {K. A. Gschneidner, Jr.}, L.~Eyring (Eds.), Handbook
on the
  Physics and Chemistry of Rare Earths, Vol.~7, Elsevier, Amsterdam, 1984.

\bibitem{Bodak97}
O.~Bodak, P.~Salamakha, O.~Sologub, J. Alloys Compd. 256 (1997)
L8.

\bibitem{Huang05b}
M.~Huang, unpublished.

\bibitem{deBoer88}
F.~R. de~Boer, R.~Boom, W.~C.~M. Mattens, A.~R. Miedema, A.~K.
Niessen, in:
  F.~R.~D. Boer, D.~G. Pettifor (Eds.), Cohesion in Metals, Transition Metal
  Alloys, Vol.~1, Elsevier, Amsterdam, 1988.

\bibitem{Myskiv74}
M.~Mys'kiv, O.~Bodak, E.~I. Gladyshevskii, in: R.~Rykhal (Ed.),
Tezisy Dokl.
  Vses. Konf. Kristallokhim. Intermet. Soedin. 2nd Ed., 1974, p.~31.

\bibitem{Janssen04}
{Y. Janssen et al.}, unpublished.

\bibitem{Rykhal71}
R.~M. Rykhal, O.~S. Zarechnyuk, N.~V. German, Russ.
Metall.-Metall.-U. 6 (1971)
  205.

\bibitem{Nakamura04}
H.~Nakamura, S.~Giri, T.~Kohara, J. Phys. Soc. Jpn. 73 (2004)
2971.

\bibitem{Thiede98}
V.~M.~T. Thiede, W.~Jeitschko, Z. Naturforsch. B 53 (1998) 673.

\bibitem{Jansson92}
A.~Jansson, Metall. Trans. A 23 (1992) 2953.

\bibitem{Buschow76}
K.~H.~J. Buschow, J.~H.~N. {van Vucht}, W.~W. {van den Hoogenhof},
J.
  Less-Common Met. 50 (1976) 145.

\end{thebibliography}
\end{document}